\documentclass[prd,twocolumn,groupedaddress,showpacs,nofootinbib]{revtex4}
\usepackage{graphicx}
\usepackage{dcolumn}
\usepackage{amssymb}
\usepackage{mathrsfs}
\usepackage{amsmath}
\usepackage{epsfig}
\usepackage[dvips]{color}
\usepackage{hhline}

\begin{document}

\title{Minimal inverse seesaw accompanied by Dirac fermionic dark matter}

\author{Pei-Hong Gu}

\email{peihong.gu@sjtu.edu.cn}

\affiliation{School of Physics and Astronomy, Shanghai Jiao Tong University, 800 Dongchuan Road, Shanghai 200240, China}

\begin{abstract}

We present a minimal inverse seesaw mechanism by resorting to a $U(1)_{B-L}^{}$ gauge symmetry. In order to cancel the gauge anomalies, we introduce seven neutral fermions among which four participate in the inverse seesaw to induce two nonzero neutrino mass eigenvalues, two forms a stable Dirac fermion to become a dark matter, while the last one keeps massless but decouples early. In this inverse seesaw, two neutral fermions are the usual right-handed neutrinos while the other two have a small Majorana mass term. An additional seesaw mechanism for generating these small Majorana masses also explains the cosmic baryon asymmetry in association with the sphaleron processes.

\end{abstract}


\maketitle

\section{Introduction}

The fact that three flavors of neutrinos should be massive and mixed has been confirmed by various neutrino oscillation experiments \cite{pdg2018}. Moreover, the neutrinos should be extremely light to satisfy the cosmological constraints \cite{pdg2018}.  The tiny but nonzero neutrino masses can be naturally explained by the famous seesaw mechanism \cite{minkowski1977,yanagida1979,grs1979,ms1980}. Some seesaw models such as the type-I  \cite{minkowski1977,yanagida1979,grs1979,ms1980}, type-II  \cite{mw1980,sv1980,cl1980,lsw1981,ms1981} and type-III \cite{flhj1989} seesaw can further accommodate a leptogenesis mechanism \cite{fy1986} to understand the cosmic baryon asymmetry \cite{pdg2018}, which is much much bigger than the value induced in the standard model (SM). In this seesaw-leptogenesis framework, additionally heavy particles couple to the SM lepton and Higgs doublets so that their decays can produce a lepton asymmetry stored in the SM leptons \cite{fy1986,lpy1986,fps1995,ms1998,bcst1999,hambye2001,di2002,gnrrs2003,hs2004,bbp2005}. In association with the sphaleron processes \cite{krs1985}, the produced lepton asymmetry can be partially converted to a baryon asymmetry. Meanwhile, the small neutrino masses can be induced by integrating out the heavy particles from the same interactions.

The inverse seesaw mechanism is another attractive seesaw scenario \cite{mv1986}. In the inverse seesaw models, several right-handed neutrinos can have a sizable mass term with the same number of other neutral fermions which are assigned to a small Majorana mass term. Due to such small Majorana masses, the right-handed neutrinos can be allowed to significantly mix with the left-handed neutrinos. The right-handed neutrinos thus may be verified experimentally. Remarkably, the small Majorana mass term of the neutral fermions is crucial to the testability of the inverse seesaw. It has been shown such small Majorana masses can be suppressed by additionally heavy fermion \cite{adefhv2018,gu2019} and/or scalar \cite{gu2019} singlets after a $U(1)_X^{}$ global symmetry is spontaneously broken. The same interactions can also account for the generation of the baryon asymmetry.

In this paper we shall develop our recent work \cite{gu2019} to solve the dark matter puzzle, which is another big challenge to the SM. Specifically, the SM $SU(3)_c^{}\times SU(2)_L^{}\times U(1)_Y^{}$ gauge symmetries will be extended by a $U(1)_{B-L}^{}$ gauge symmetry. In order to cancel the gauge anomalies, we introduce seven neutral fermions including two usual right-handed neutrinos. After a Higgs singlet develops its vacuum expectation value (VEV) for spontaneously breaking the $U(1)_{B-L}^{}$ symmetry, two neutral fermions can obtain their small Majorana masses by integrating out some heavy scalar or fermion singlets, meanwhile, they can have a sizable mass term with the right-handed neutrinos. After the SM Higgs doublet drives the electroweak symmetry breaking, the two right-handed neutrinos and the three left-handed neutrinos can acquire their Dirac masses. We thus can realize an inverse seesaw to give a rank-2 neutrino mass matrix with two nonzero eigenvalues. Within this framework, the heavy scalar or fermion singlet decays can explain the cosmic baryon asymmetry in association with the sphaleron processes. On the other hand, another two neutral fermions can form a Dirac particle due to their Yukawa coupling with the $U(1)_{B-L}^{}$ Higgs singlet. This Dirac fermion can serve as a stable dark matter particle. As for the seventh neutral fermion, it keeps massless but decouples early.

\section{Fermions and scalars}

The SM fermions now are gauged by a $U(1)_{B-L}^{}$ symmetry besides the $SU(3)_c^{} \times SU(2)^{}_{L}\times U(1)_Y^{}$ groups, 
\begin{eqnarray}
&&\begin{array}{l}q^{}_{L}(3,2,+\frac{1}{6},+\frac{1}{3}),\end{array} \!\!
\begin{array}{l}d^{}_{R}(3,1,-\frac{1}{3},+\frac{1}{3}),\end{array}  \!\!
 \begin{array}{l}u^{}_{R}(3,1,+\frac{2}{3},+\frac{1}{3});\end{array} \nonumber\\
[2mm]
&&\begin{array}{l}l^{}_{L}(1,2,-\frac{1}{2},-1),\end{array} \!\!
\begin{array}{l}e^{}_{R}(1,1,-1,-1).\end{array} 
\end{eqnarray}
Here and thereafter the brackets following the fields describe the transformations under the $SU(3)_c^{} \times SU(2)^{}_{L}\times U(1)_Y^{}\times U(1)_{B-L}^{}$ gauge groups. For simplicity, we do not show the indices of the three generations of fermions. In order to cancel the gauge anomalies, we can introduce some neutral fermions with appropriate $U(1)_{B-L}^{}$ charges \cite{mp2007,pry2016,gu2019-2}. In the present work, we take seven neutral fermions including two usual right-handed neutrinos,
\begin{eqnarray}
&&\begin{array}{l}
\nu^{}_{R1,2}(1,1,0,-1),\end{array}  \begin{array}{l}S^{}_{R1,2}(1,1,0,+\frac{1}{2}),\end{array} \nonumber\\
&& \begin{array}{l}\chi^{}_{R1}(1,1,0,\frac{1+\sqrt{153}}{4}),\end{array}  \begin{array}{l}\chi^{}_{R2}(1,1,0,\frac{1-\sqrt{153}}{4}),\end{array} \nonumber\\
&&\begin{array}{l}
\zeta^{}_{R}(1,1,0,-\frac{5}{2}).\end{array}\end{eqnarray}

While the SM Higgs doublet, 
\begin{eqnarray}
\begin{array}{l}\phi^{}(1,2,-\frac{1}{2},0),\end{array}
\end{eqnarray}
is responsible for the electroweak symmetry breaking as usual, we introduce a Higgs singlet, 
\begin{eqnarray}
\begin{array}{l}
\xi(1,1,0,+\frac{1}{2}),\end{array}
\end{eqnarray}
to spontaneously break the $U(1)_{B-L}^{}$ symmetry. Furthermore, our model contains some heavy SM-singlet scalars or fermions,
\begin{eqnarray}
\begin{array}{l}X^{}_{Ra}(1,1,0,0)\,,~~\Sigma_b^{}(1,1,0,+1)\,,~~(a,b=1,...)\,.\end{array}
\end{eqnarray}

For simplicity, we do not write down the full Yukawa and scalar interactions. Indeed, we will show later the inverse seesaw and the leptogenesis only depend on the following terms, 
\begin{eqnarray}
\label{lag}
\mathcal{L}&\supset& \mathcal{L}_{\textrm{I}/\textrm{II}}^{}- \left(y \bar{l}_{L}^{} \phi \nu_R^{} + f \xi \bar{\nu}_R^{c} S_R^{} +\textrm{H.c.}\right)~~\textrm{with}\nonumber\\
&&\mathcal{L}_{\textrm{I}}\supset -\frac{1}{2}M_X^{} \bar{X}_R^{}X_R^{c} -g_X^{} \xi \bar{S}_R^{} X_R^{c} +\textrm{H.c.}\,,\nonumber\\
&& 
\mathcal{L}_{\textrm{II}}^{}\supset -M_\Sigma^2 \Sigma^\dagger_{}\Sigma + \rho_\Sigma^{}\Sigma \xi \xi+ \frac{1}{2}g_\Sigma^{}\Sigma \bar{S}_R^c S_R^{} +\textrm{H.c.}\,.\nonumber\\
&&
\end{eqnarray}
Without loss of generality and for convenience, we can take the mass matrices $M_\Sigma^{}$ and $M_{X}^{}$ to be real and diagonal. Accordingly, we can define the Majorana fermions as below,
\begin{eqnarray}
X_a^{}= X_{Ra}^{} + X_{Ra}^c = X^c_{a}\,.
\end{eqnarray}
Note if the heavy scalar singlet(s) $\Sigma_b^{}$ and the heavy fermion singlet(s) $X_{Ra}^{}$ are both introduced to the models, they should have the Yukawa interactions of the form $\Sigma \bar{\nu}_{R}^{c}X_R^{} +\textrm{H.c.}$. However, such terms are harmful for the realisation of an inverse seesaw. So, we will not consider this case. Instead, we will focus on the case with either the heavy scalar singlet(s) $\Sigma_b^{}$ or the heavy fermion singlet(s) $X_{Ra}^{}$.

We then move to the other neutral fermions $\chi_{R1,2}^{}$ and $\zeta_R^{}$. We can easily find that because of the $U(1)_{B-L}^{}$ symmetry, the $\zeta_R^{}$ fermion is forbidden to have any Yukawa coupling, but the $\chi_{R1,2}^{}$ fermions are allowed to have a Yukawa coupling as below,
\begin{eqnarray}
\label{dm}
\mathcal{L}&\supset& - y_\chi^{} \left(\xi \bar{\chi}_{R1}^{} \chi_{R2}^c +\textrm{H.c.}\right)\,.
\end{eqnarray}
We will show in the following that the $\chi_{R1,2}^{}$ fermions provide a dark matter candidate while the $\zeta_R^{}$ fermion keeps massless and decouples safely.

\section{Minimal inverse seesaw}

\begin{figure*}
\vspace{8cm} \epsfig{file=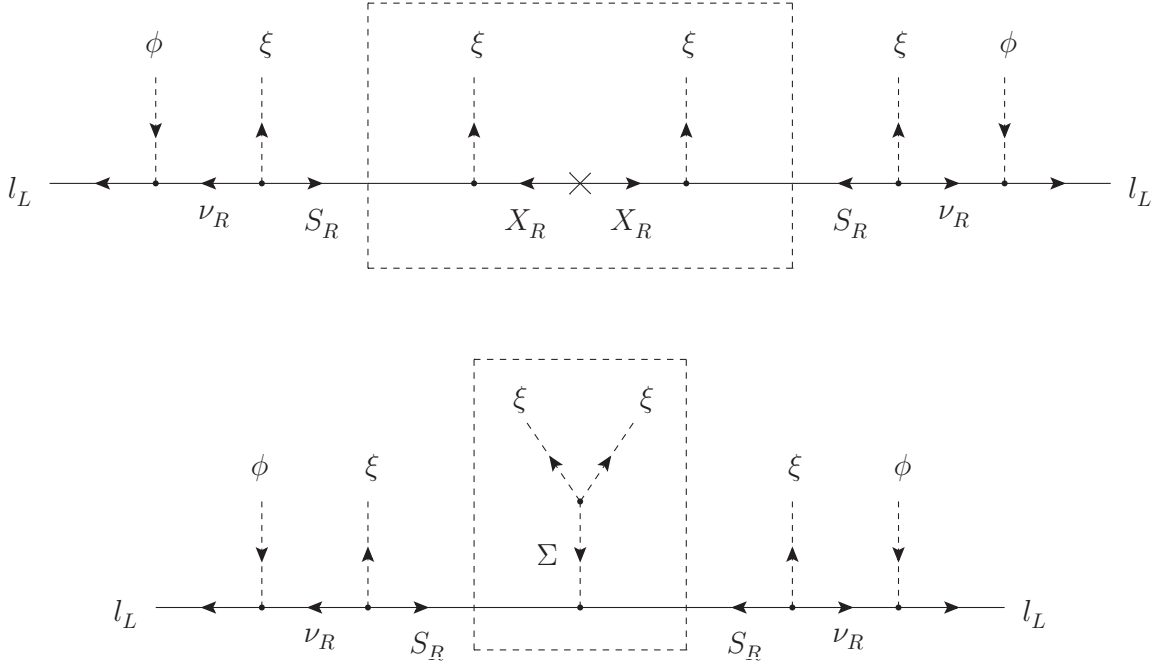, bbllx=7.5cm, bblly=6.0cm,
bburx=17.5cm, bbury=16cm, width=8cm, height=8cm, angle=0,
clip=0} \vspace{-6cm} \caption{\label{numass} The inverse seesaw mechnasim for generating the tiny neutrino masses. The interactions in the boxes provide an additional seesaw mechanism for generating a small Majorana mass term which is crucial to the inverse seesaw mechanism. }
\end{figure*}

When the Higgs singlet $\xi$ acquires its VEV for the spontaneous $U(1)_{B-L}^{}$ symmetry breaking, the two neutral fermions $S_{R1,2}^{}$ can obtain a Majorana mass term by integrating out the Majorana fermions $X_a^{}$ or the Higgs singlets $\Sigma_b^{}$ in Eq. (\ref{lag}), i.e. 
\begin{eqnarray}
\mathcal{L}&\supset&- \frac{1}{2}\sum_{i,j=1,2}^{}\left(\mu_S^{}\right)_{ij}^{} \bar{S}_{Ri}^c S_{Rj}^{} +\textrm{H.c.}~~\textrm{with}\nonumber\\
&&\left(\mu_S^{}\right)_{ij}^{}=- \sum_{b=1}^{n\geq 2}\left(g_{X}^{\ast}\right)_{ib}^{} \frac{ \langle\xi\rangle^2_{}}{M_{X_b}^{}} \left(g_{X}^{\dagger}\right)_{bj}^{}~~\textrm{or}\nonumber\\
&&\left(\mu_S^{}\right)_{ij}^{} = - \sum_{a=1}^{n\geq 2}\left(g_{\Sigma_{a}}^{} \right)_{ij}^{}\frac{\rho_{\Sigma_a}^{} \langle\xi\rangle^2_{}}{M_{\Sigma_a}^2}\,.
\end{eqnarray}
Clearly, as the Majorana fermions $X_a^{}$ or the Higgs singlets $\Sigma_b^{}$ are assumed to be much heavier than the $U(1)_{B-L}^{}$ breaking scale $\langle\xi\rangle$, the above Majorana masses $\mu_S^{}$ can be highly suppressed in a natural way. This Majorana mass generation is similar to the conventional type-I or type-II seesaw mechanism. 

At this stage, the two neutral fermions $S_{R1,2}^{}$ can also mix with the two right-handed neutrinos $\nu_{R1,2}^{}$ to form the quasi-Dirac fermions, i.e.
\begin{eqnarray}
\mathcal{L}&\supset& -\sum_{i,j=1,2}^{}(m_N^{})_{ij} \bar{N}_{Ri}^{} N_{Lj}^{} +\textrm{H.c.}^{} ~~\textrm{with}\nonumber\\
&&N=\nu_R^{}+S_R^c\,,~~m_N^{} = f\langle\xi\rangle\,.
\end{eqnarray}
Subsequently, the Higgs doublet $\phi$ develops its VEV for the electroweak symmetry breaking. The two right-handed neutrinos $\nu_{R1,2}^{}$ can further have a Dirac mass term with and the three left-handed neutrinos $\nu_{L\alpha}^{}(\alpha=e,\mu,\tau)$, i.e.
 \begin{eqnarray}
\mathcal{L}&\supset& -\sum_{\alpha=e,\mu,\tau \atop i=1,2}^{} (m_D^{})_{\alpha i} \bar{\nu}_{L\alpha}^{} \nu_{Ri}^{} +\textrm{H.c.}^{} ~~\textrm{with}~~m_D^{} = y\langle\phi\rangle\,. \nonumber\\
&&
\end{eqnarray}

In the limiting case,
 \begin{eqnarray}
m_N^{}\gg m_D^{}\,,~\mu_S^{}\,,
\end{eqnarray}
the left-handed neutrinos $\nu_L^{}$ can obtain a tiny Majorana mass term, 
 \begin{eqnarray}
 \label{num}
\mathcal{L}&\supset& -\frac{1}{2}m_\nu^{} \bar{\nu}_L^{} \nu_L^c +\textrm{H.c.}~~\textrm{with}\nonumber\\
&&m_\nu^{}=m_D^{} \frac{1}{m_N^\dagger}\mu_S^{}\frac{1}{m_N^{\ast}}m_D^{T}\,.
\end{eqnarray}
This is an application of the so-called inverse seesaw mechanism.

The neutrino masses (\ref{num}) can be understood in Fig. \ref{numass}. It should be noted that only two right-handed neutrinos $\nu_{R1,2}^{}$ and the same number of fermion singlets $S_{R1,2}^{}$ participate in the neutrino mass generation. Such rank-2 neutrino mass matrix can only have two nonzero eigenvalues. In this sense, we may refer to our model as a minimal inverse seesaw \cite{sv1980}.

\section{Baryon asymmetry}

In this section we shall demonstrate how to generate the baryon asymmetry in our models.

\subsection{The heavy fermion or Higgs singlet decays}

\begin{figure*}
\vspace{7.5cm} \epsfig{file=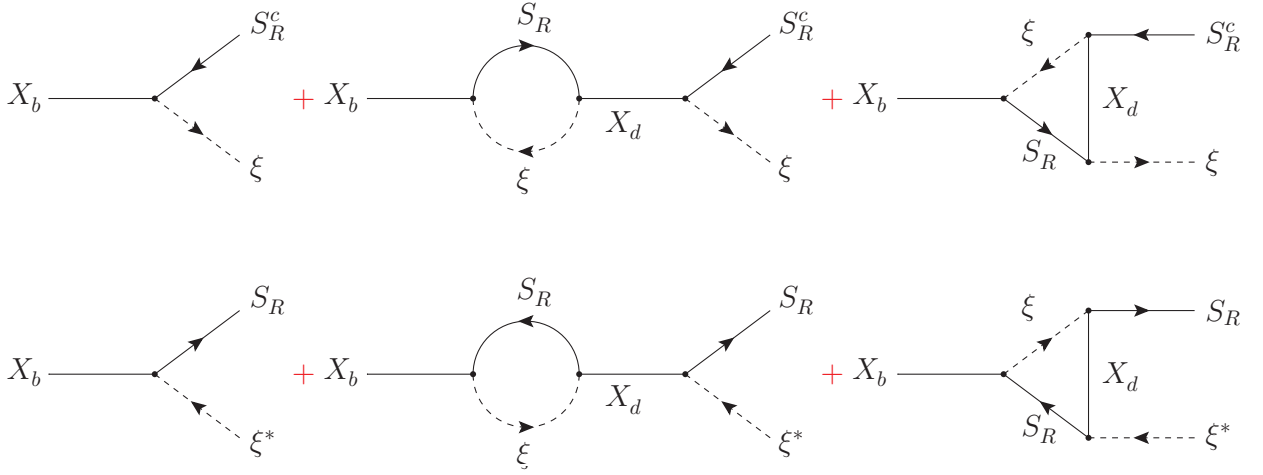, bbllx=8cm, bblly=6.0cm,
bburx=18cm, bbury=16cm, width=8cm, height=8cm, angle=0,
clip=0} \vspace{-8.5cm} \caption{\label{fdecay} The heavy fermion singlet decays.}
\end{figure*}

As shown in  Fig. \ref{fdecay}, the heavy fermion singlet $X_b^{}$ have two decay modes, 
\begin{eqnarray}
X_b^{} \rightarrow S_R^{c}+\xi\,,~~ X_b^{} \rightarrow S_R^{}+ \xi^\ast_{} \,.
\end{eqnarray}
As long as the CP is not conserved, we can expect a CP asymmetry in the above decays,
\begin{eqnarray}
\varepsilon_{X_b}^{}&=& \frac{\Gamma(X_b^{} \rightarrow S_R^c+\xi)- \Gamma(X_b^{} \rightarrow S_R^{}+\xi^\ast_{})}{\Gamma_{X_b^{}}^{}}\,,
\end{eqnarray}
with $\Gamma_{\Sigma_a}^{}$ being the total decay width,
\begin{eqnarray}
\Gamma_{X_b}^{}&=& \Gamma(X_b^{} \rightarrow S_R^{c}+\xi)+\Gamma( X_b^{} \rightarrow S_R^{}+ \xi^\ast_{})\,.
\end{eqnarray}
We can calculate the decay width at tree level and the CP asymmetry at one-loop order, i.e.
\begin{eqnarray}
\Gamma_{X_b}^{}&=&\frac{1}{16\pi} \left(g_X^\dagger g_X^{}\right)_{bb}^{}M_{X_b}^{}\\
[2mm]
\varepsilon_{X_b}^{}&=&-\frac{1}{16\pi} \sum_{d\neq b}^{}\frac{\textrm{Im}\left\{\left[\left(g_{X}^T g_{X}^{\ast}\right)_{bd}\right]^2_{}\right\}}{\left(g_X^\dagger g_X^{}\right)_{bb}^{}}\left\{\frac{M_{X_b}^{}M_{X_d}^{}}{M_{X_d}^2- M_{X_b}^2}\right.\nonumber\\
&&\left.+\frac{2M_{X_d}^{}}{M_{X_b}^{}}\left[1-\left(1+\frac{M_{X_d}^{2}}{M_{X_b}^{2}}\right)\ln\left(1+\frac{M_{X_b}^{2}}{M_{X_d}^{2}}\right)\right]\right\}\,.\nonumber\\
&&
\end{eqnarray}
For a nonzero CP asymmetry $\varepsilon_{X_b}^{}$, we need at least two heavy fermion singlets $X_{1,...,n\geq 2}^{}$.

\begin{figure*}
\vspace{9cm} \epsfig{file=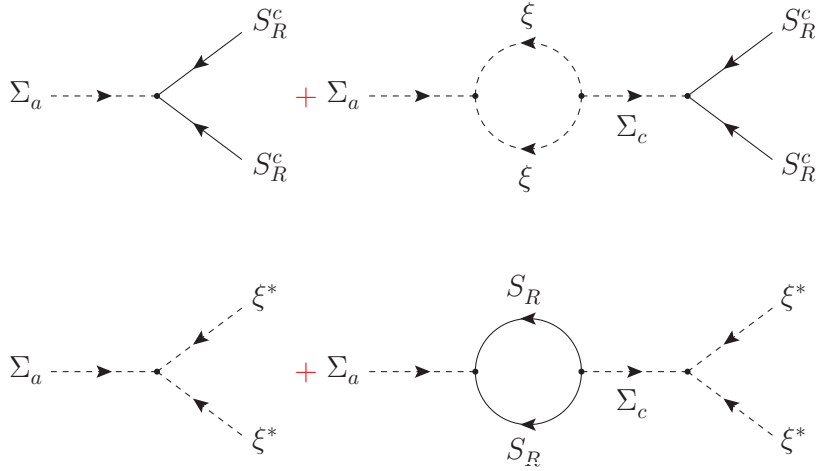, bbllx=5.5cm, bblly=6.0cm,
bburx=15.5cm, bbury=16cm, width=8cm, height=8cm, angle=0,
clip=0} \vspace{-8.5cm} \caption{\label{sdecay} The heavy Higgs singlet decays.}
\end{figure*}

As for the heavy Higgs singlet $\Sigma_a^{}$, their decay modes are 
\begin{eqnarray}
\Sigma_a^{} \rightarrow  S_R^{c}+S_R^{c}\,,~~ \Sigma_a^{} \rightarrow \xi^\ast_{}+\xi^\ast_{} \,.
\end{eqnarray}
The relevant diagrams are shown in Fig. \ref{sdecay}. The tree-level decay width and the one-loop CP asymmetry can be calculated by
\begin{eqnarray}
\Gamma_{\Sigma_a}^{}&=&\Gamma(\Sigma_a^{} \rightarrow S_R^{c}+S_R^{c})  + \Gamma( \Sigma_a^{} \rightarrow \xi^\ast_{}+\xi^\ast_{})\nonumber\\
&=&\Gamma( \Sigma_a^{\ast} \rightarrow  S_R^{}+S_R^{}  )+\Gamma( \Sigma^\ast_{a} \rightarrow \xi+\xi)\nonumber\\
&=&\frac{1}{8\pi}\left[\textrm{Tr}\left(g_{\Sigma_a^{}}^\dagger g_{\Sigma_a}^{}\right)+\frac{\rho_{\Sigma_a}^2}{M_{\Sigma_b}^2}\right]M_{\Sigma_a}^{}\,,
\end{eqnarray}
\begin{eqnarray}
\varepsilon_{\Sigma_a}^{}&=&2 \frac{\Gamma(\Sigma_a^{} \rightarrow S_R^{c}+S_R^{c}  )-\Gamma( \Sigma_a^{\ast} \rightarrow  S_R^{}+S_R^{}  )}{\Gamma_{\Sigma_a^{}}^{}}\nonumber\\
&=&2\frac{\Gamma( \Sigma^\ast_{a} \rightarrow \xi+\xi)-\Gamma( \Sigma_a^{} \rightarrow \xi^\ast_{}+\xi^\ast_{})}{\Gamma_{\Sigma_a^{}}^{}}\nonumber\\
&=& -\frac{1}{\pi}\sum_{c\neq a}^{}\frac{\textrm{Im}\left[\textrm{Tr}\left(g_{\Sigma_a}^\dagger g_{\Sigma_c}^{}\right)\rho_{\Sigma_a}^{}\rho_{\Sigma_c}^{}\right]}{\textrm{Tr}\left(g^\dagger_{\Sigma_a} g_{\Sigma_a}^{}\right)+ \frac{\rho_{\Sigma_a}^2}{M_{\Sigma_a}^{2}}}\frac{1}{M_{\Sigma_c}^2-M_{\Sigma_a}^2}\,.\nonumber\\
&&
\end{eqnarray}
A nonzero CP asymmetry $\varepsilon_{\Sigma_a}^{}$ requires the existence of at least two heavy Higgs singlets $\Sigma_{1,...,n\geq 2}^{}$.

After the heavy fermion singlets $X_b^{}$ or the heavy Higgs singlets $\Sigma_a^{}$ go out of equilibrium, their decays can generate an asymmetry $A_S^{}$ stored in the fermion singlets $S_R^{}$. For demonstration, we can simply assume the lightest heavy fermion singlet $X_1^{}$ or the lightest heavy Higgs singlet $\Sigma_1^{}$ to be much lighter than the other heavy Higgs or fermion singlets. The $A_S^{}$ asymmetry then should mainly come from the $X_1^{}$ or $\Sigma_1^{}$  decays, i.e.
\begin{eqnarray}
\label{xasymmetry}
A_S^{}&=& \varepsilon_{X_1/\Sigma_1 }^{}\left(\frac{n^{eq}_{\Sigma_1/X_1} }{s}\right)\left|_{T=T_D^{}}^{}\right.,
\end{eqnarray}
where the symbols $n^{eq}_{X_1/\Sigma_1}$ and $T_D^{}$ respectively are the equilibrium number density and the decoupled temperature of the lightest heavy Higgs or fermion singlets, while the character $s$ is the entropy density of the universe. In this case, the CP asymmetries $\varepsilon_{X_1/\Sigma_1}^{}$ can be simplified by 
\begin{eqnarray}
\varepsilon_{X_1}^{}&\simeq & \frac{1}{8\pi}\frac{\textrm{Im}\left[\left(g_{X}^T \mu_S^{} g_{X}^{}\right)_{11}\right] M_{X_1}^{}}{\left(g_X^\dagger g_X^{}\right)_{11}^{}\langle\xi\rangle^2_{}}\nonumber\\
&\lesssim& \frac{1}{8\pi}\frac{\mu_{\textrm{max}}^{} M_{X_1}^{}}{\langle\xi\rangle^2_{}} \,,
\end{eqnarray}
\begin{eqnarray}
\varepsilon_{\Sigma_1}^{}&\simeq & \frac{1}{\pi}\frac{\textrm{Im}\left[\textrm{Tr}\left(g_{\Sigma_1}^\dagger \mu_S^{} \right)\right] \rho_{\Sigma_1}^{}}{\left[\textrm{Tr}\left(g^\dagger_{\Sigma_1} g_{\Sigma_1}^{}\right)+ \frac{\rho_{\Sigma_1^{}}^2}{M_{\Sigma_1}^{2}} \right] \langle\xi\rangle^2_{} }\nonumber\\
&\leq&  \frac{1}{\pi}\frac{\textrm{Im}\left[\textrm{Tr}\left(g_{\Sigma_1}^\dagger \mu_S^{} \right)\right] \rho_{\Sigma_1}^{}}{2\sqrt{\textrm{Tr}\left(g^\dagger_{\Sigma_1} g_{\Sigma_1}^{}\right)\frac{\rho_{\Sigma_1}^2}{M_{\Sigma_1}^{2}} } \langle\xi\rangle^2_{} }\nonumber\\
&\lesssim&   \frac{1}{2\pi}\frac{ \mu_{\textrm{max}}^{} M_{\Sigma_1}^{}}{ \langle\xi\rangle^2_{} }\,,
\end{eqnarray}
with $\mu_{\textrm{max}}^{} $ being the largest eigenvalue of the Majorana mass matrix $\mu_S^{}$.

Note the Majorana masses $\mu_S^{}$ can lead to an $S_R^{}-S_R^c$ oscillation, which tends to wash out the $A_S^{}$ asymmetry. However, this oscillation will not go into equilibrium before the sphalerons stop working, i.e.\begin{eqnarray}
\label{ssbar}
&&\left[\Gamma_{S_R^{}-S_R^c}^{}>H(T)\right]\left|_{T\ll  100\,\textrm{GeV}}^{}\right.\nonumber\\
&&\textrm{with}~~\Gamma_{S_R^{}-S_R^c}^{}\sim\left\{\begin{array}{cc} \frac{\mu_S^2}{T}&\textrm{for}~~T>m_N^{}\,,\\
[2mm]
\frac{\mu_S^2}{m_N^{}}&\textrm{for}~~T<m_N^{}\,.\end{array}\right.
\end{eqnarray}
Here $H(T)$ is the Hubble constant to be given later. This means we need not take the Majorana masses $\mu_S^{}$ into account before the electroweak symmetry breaking.

\subsection{The lepton-to-baryon conversion}

The asymmetry $A_S^{}$ stored in the fermion singlets $S_R^{}$ will lead to a lepton asymmetry stored in the SM lepton doublets $l_L^{}$ because of the related Yukawa interactions in Eq. (\ref{lag}). The sphaleron processes then can partially transfer this lepton asymmetry to a baryon asymmetry. We can impose a global symmetry of lepton number under which the right-handed neutrinos $\nu_R^{}$ and the fermion singlets $S_{R}^{}$ carry the lepton numbers $+1$ and $-1$, respectively. The global lepton number in our models (\ref{lag}) then is softly broken by the Majorana masses of the heavy fermion singlets $X_R^{}$ or the cubic couplings of the heavy Higgs singlets $\Sigma$ to the $U(1)_{B-L}^{}$ Higgs singlet $\xi$. For such global lepton number, the $A_S^{}$ asymmetry equals to a lepton asymmetry $L_S^{}$ stored in the fermion singlets $S_R^{}$. The lepton-to-baryon conversion thus will not be affected by the $U(1)_{B-L}^{}$ gauge symmetry breaking. Therefore, we can conveniently analysize the lepton-to-baryon conversion after the $U(1)_{B-L}^{}$ symmetry breaking. For this purpose, we denote $\mu_{q}^{}$, $\mu_d^{}$, $\mu_u^{}$, $\mu_l^{}$, $\mu_e^{}$, $\mu_\nu^{}$, $\mu_S^{}$ and $\mu_\phi^{}$ for the chemical potentials of the fields $q_{L}^{}$, $d_{R}^{}$, $u_{R}^{}$, $l_L^{}$, $e_R^{}$, $\nu_R^{}$, $S_R^{}$ and $\phi$. The SM Yukawa interactions are in equilibrium and hence yield \cite{ht1990},
\begin{eqnarray}
\label{chemical1}
-\mu_{q}^{}+\mu_{d}^{}-\mu_{\phi}^{}&=&0\,,\\
\label{chemical2}
 -\mu_{q}^{}+\mu_{u}^{}+\mu_{\phi}^{}&=&0\,,\\
 \label{chemical3} 
 -\mu_{l}^{}+\mu_{e}^{}-\mu_{\phi}^{}&=&0\,,\\
 -\mu_{l}^{}+\mu_{\nu}^{} + \mu_\phi^{}&=&0\,,
\end{eqnarray}
the fast sphalerons constrain \cite{ht1990},
\begin{eqnarray}
\label{chemical4}
3\mu_{q}^{}+\mu_{l}^{}&=&0\,,
\end{eqnarray}
while the neutral hypercharge in the universe requires \cite{ht1990},
\begin{eqnarray}
\label{chemical5}
3\left( \mu_{q}^{} -\mu_{d}^{}+2\mu_{u}^{}-\mu_{l}^{} -\mu_{e}^{}\right)-2\mu_{\phi}^{} =0\,.
\end{eqnarray}
In addition, the mass term between the right-handed neutrinos $\nu_R^{}$ and the fermion singlets $S_R^{}$ are also in equilibrium. This means
\begin{eqnarray}
\label{chemical6}
\mu_S^{}+\mu_\nu^{}&=&0\,.
\end{eqnarray}
In the above Eqs. (\ref{chemical1}-\ref{chemical6}), we have identified the chemical potentials of the different-generation fermions because the Yukawa interactions establish an equilibrium between the different generations. By solving Eqs. (\ref{chemical1}-\ref{chemical6}), we can express the chemical potentials by
\begin{eqnarray}
\!\!\!\!&&\mu_\phi^{}=-\frac{4}{7}\mu_l^{}\,,~~\mu_q^{}=-\frac{1}{3}\mu_l^{}\,,~~\mu_d^{}=-\frac{19}{21}\mu_l^{}\,,~~\mu_u^{}=\frac{5}{21}\mu_l^{}\,,\nonumber\\
\!\!\!\!&&\mu_{e}^{}=\frac{3}{7}\mu_l^{}\,,~~\mu_{\nu}^{}=\frac{11}{7}\mu_l^{}\,,~~\mu_S^{}=-\frac{11}{7}\mu_l^{}\,.
\end{eqnarray}

Now the global baryon number can be given by
\begin{eqnarray}
B= 3(2\mu_q^{} + \mu_d^{} + \mu_u^{}) = -4 \mu_l^{}\,.
\end{eqnarray}
As for the global lepton number, it should be 
\begin{eqnarray}
\label{lll}
L= 3(2\mu_L^{}+\mu_e^{}) + 2(\mu_\nu^{}-\mu_S^{}) = \frac{95}{7}\mu_l^{}\,.
\end{eqnarray}
The baryon and lepton numbers then can be transferred from a conserved $B-L$ number, i.e.  
\begin{eqnarray}
\label{bl1}
B=\frac{28}{123}(B-L)\,,~~L=-\frac{95}{123}(B-L)\,.
\end{eqnarray}
If the quasi-Dirac fermions $N=\nu_R^{}+S_R^c$ keep relativistic before the electroweak symmetry breaking, the final baryon number can be given by (\ref{bl1}). Alternatively, the $N$ fermions have become non-relativistic and have completely decayed before the electroweak symmetry breaking. In this case, the global lepton number (\ref{lll}) should be modified as 
\begin{eqnarray}
L= 3(2\mu_L^{}+\mu_e^{}) = \frac{51}{7}\mu_l^{}\,,
\end{eqnarray}
and hence the baryon and lepton numbers from the conserved $B-L$ number should be   
\begin{eqnarray}
\label{bl2}
B=\frac{28}{79}(B-L)\,,~~L=-\frac{51}{79}(B-L)\,.
\end{eqnarray}

\subsection{A numerical example}

For simplicity, we consider the weak washout condition \cite{kt1990},
\begin{eqnarray}
\label{weak}
\!\!\!\!\left.\left[\Gamma_{\Sigma_1/X_1}^{}< H(T)=\left[\frac{8\pi^{3}_{}g_{\ast}^{}(T)}{90}\right]^{\frac{1}{2}}_{}\frac{T^2_{}}{M_{\textrm{Pl}}^{}}\right]\right|_{T=M_{\Sigma_1^{}/X_1^{}}}^{}\,.
\end{eqnarray}
The final baryon asymmetry then can approximate to \cite{kt1990}
\begin{eqnarray}
\label{bauf4}
B^f_{}\sim c \frac{\varepsilon_{\Sigma_1/X_1}^{}}{g_\ast^{}}~~\textrm{with}~~c=-\frac{28}{79}~\textrm{or}~-\frac{28}{123}\,.
\end{eqnarray}
Here $H(T)$ is the Hubble constant with $M_{\textrm{Pl}}^{}\simeq 1.22\times 10^{19}_{}\,\textrm{GeV}$ being the Planck mass and $g_{\ast}^{}(T)=123$ being the relativistic degrees of freedom (the SM fields plus the seven neutral fermions ($\nu_{R1,2}^{},S_{R1,2}^{},\chi_{R1,2}^{},\zeta_R^{}$), the $U(1)_{B-L}^{}$ Higgs singlet $\xi$ and the $U(1)_{B-L}^{}$ gauge boson.).

As an example, we choose 
\begin{eqnarray}
\langle\xi\rangle= 40\,\textrm{TeV}\,,
\end{eqnarray}
and then take
\begin{eqnarray}
\!\!\!\!\!\!\!\!\!\!\!\!&&M_{\Sigma_1^{}}^{}=10^{14}_{}\,\textrm{GeV}\,,~\rho_{\Sigma_1^{}}^{}=3\times 10^{12}_{}\,\textrm{GeV}\,, ~g_{\Sigma_1^{}}^{}= \mathcal{O}(0.03)\,; \nonumber\\
\!\!\!\!\!\!\!\!\!\!\!\!&&\textrm{or}~~M_{X_1^{}}^{}=10^{14}_{}\,\textrm{GeV}\,,~g_{X}^{}= \mathcal{O}(0.03)\,.
\end{eqnarray}
With these inputting, we can have
\begin{eqnarray}
\label{output1}
\mu_S^{}= \mathcal{O}(10\,\textrm{eV})\,,
\end{eqnarray}
as well as 
\begin{eqnarray}
\frac{\Gamma_{\Sigma_1^{}/X_1^{}}^{}}{H(T)}\left|_{T=M_{\Sigma_1^{}/X_1^{}}^{}}^{}\right.=\mathcal{O}(0.1)\,,~~\varepsilon_{\Sigma_1^{}/X_1^{}}^{\textrm{max}}=\mathcal{O}( 10^{-5}) \,.
\end{eqnarray}
So, the final baryon asymmetry (\ref{bauf4}) can arrive at the observed value $B^f_{}\sim 10^{-10}_{}$.

We further take 
\begin{eqnarray}
f\sim \mathcal{O}(10^{-3}_{}-0.1)\,,~y\sim \mathcal{O}(10^{-2}_{}-1)\,,
\end{eqnarray}
to give  
\begin{eqnarray}
\label{output2}
\!\!m_N^{} \sim \mathcal{O}(10-1000\,\textrm{GeV})\,, ~~m_D^{}\sim \mathcal{O}(1-100\,\textrm{GeV})\,.
\end{eqnarray}
By inserting the outputs (\ref{output1}) and (\ref{output2}) into the inverse seesaw (\ref{num}), we can obtain the neutrino masses $m_\nu^{}=\mathcal{O}(0.1\,\textrm{eV})$.

For the above parameter choice, we also check the $S_R^{}-S_R^c$ oscillation (\ref{ssbar}) induced by the Majorana masses $\mu_S^{}$. We find this oscillation can not go into equilibrium before the electroweak symmetry breaking and hence it will not affect the production of the baryon asymmetry.

\section{Dark fermions}

As shown in Eq. (\ref{dm}), the two neutral fermions $\chi_{R1,2}^{}$ have a Yukawa coupling with the $U(1)_{B-L}^{}$ Higgs singlet $\xi$ so that they can form a Dirac particle \cite{pry2016,gu2019-2,gu2019-3} after the $U(1)_{B-L}^{}$ symmetry breaking, i.e.
\begin{eqnarray}
\mathcal{L} &\supset&  i \bar{\chi}\gamma^\mu_{}\partial_\mu^{} \chi - m_\chi^{} \bar{\chi}\chi \nonumber\\
&& \textrm{with}~~ \chi = \chi_{R1}^{}+\chi_{R2}^{c}\,,~~ m_\chi^{}= y_{\chi}^{} \langle\xi\rangle\,.
 \end{eqnarray}
Meanwhile, the another neutral fermion $\zeta_R^{}$ has no Yukawa couplings so that it should be massless.

Now the Dirac fermion $\chi$ is stable and hence leaves a dark matter relic density. The dark matter annihilation and scattering could be determined by the gauge interactions,
\begin{eqnarray}
\mathcal{L} &\supset &g_{B-L}^{} Z_{B-L}^\mu \left[\sum_{i=1}^{3}\left(\frac{1}{3}\bar{d}_i^{}\gamma_\mu^{} d_i^{} + \frac{1}{3}\bar{u}_{i}^{}\gamma_\mu^{}u_{i}^{}-\bar{e}_{i}^{}\gamma_\mu^{} e_{i}^{}  \right.\right.\nonumber\\
&&\left.\left.-\bar{\nu}_{Li}^{}\gamma_\mu^{}\nu_{Li}^{}\right)-\bar{\nu}_{R1}^{}\gamma_\mu^{}\nu_{R1}^{} -\bar{\nu}_{R1}^{}\gamma_\mu^{}\nu_{R1}^{} \right.\nonumber\\
&&\left.+\frac{1}{2}\bar{S}_{R1}^{}\gamma_\mu^{} S_{R1}^{}+\frac{1}{2}\bar{S}_{R1}^{}\gamma_\mu^{} S_{R1}^{}-\frac{5}{2} \bar{\zeta}_{R}^{}\gamma_\mu^{}\zeta_{R}^{} \right.\nonumber\\
&&\left. + \frac{1}{4} \bar{\chi}\gamma_\mu^{}\left(\sqrt{153}+\gamma_5^{}\right)\chi \right]\,.
\end{eqnarray}
The gauge coupling $g_{B-L}^{}$ then should have an upper bound from the perturbation requirement, i.e.
\begin{eqnarray}
\label{gbl}
\frac{\sqrt{153}}{4}g_{B-L}^{} < \sqrt{4\pi} \Rightarrow g_{B-L}^{}< \sqrt{ \frac{64\,\pi}{153}}\,,
\end{eqnarray} 
while the gauge boson mass $M_{Z_{B-L}}^{}$ should be
\begin{eqnarray}
M_{Z_{B-L}^{}}^{}&= &\frac{1}{ \sqrt{2}} g_{B-L}^{} \langle \xi\rangle \,.
\end{eqnarray}
Currently the experimental constraints on the $U(1)_{B-L}^{}$ symmetry breaking is \cite{afpr2017,klq2016}
\begin{eqnarray}
\label{low}
\frac{M_{Z_{B-L}^{}}^{}}{g_{B-L}^{} }  \gtrsim 7 \,\textrm{TeV} \Rightarrow \langle \xi \rangle \gtrsim 10 \,\textrm{TeV}\,.
\end{eqnarray}

The thermally averaging dark matter annihilating cross section can be given by \cite{bhkk2009}
\begin{eqnarray}
\label{ann}
\langle\sigma_{\textrm{A}}^{} v_{\textrm{rel}}^{} \rangle&=& \sum_{f=d,u,e,\nu_{L}^{},\nu_R^{},S_R^{},\zeta_R^{}}^{}\langle\sigma(\chi+\chi^c_{}\rightarrow f+f^c_{}) v_{\textrm{rel}}^{}\rangle \nonumber\\
&\simeq &\frac{ 13311 g_{B-L}^4}{128\pi} \frac{m_\chi^2}{M_{Z_{B-L}^{}}^4} \nonumber\\
&=& \frac{13311}{32\pi} \frac{m_\chi^2}{\langle\xi\rangle^4_{}} =  \frac{13311}{32\pi} \frac{y_{\chi}^2}{\langle\xi\rangle^2_{}}  \,,
\end{eqnarray}
where we have assumed
\begin{eqnarray}
\label{y34}
4m_\chi^2 \ll M_{Z_{B-L}^{}}^2 &\Rightarrow & y_{\chi}^{2}\ll \frac{1}{8} g_{B-L}^{2} < \frac{8\pi}{153}\nonumber\\
&\Rightarrow& y_\chi^{}< \sqrt{\frac{8\pi}{153}}\,.
\end{eqnarray}
The dark matter relic density then can approximate to \cite{pdg2018}
\begin{eqnarray}
\label{relic}
\Omega_\chi^{} h^2 \simeq \frac{0.1\,\textrm{pb}}{\langle \sigma_{\textrm{A}}^{} v_{\textrm{rel}}^{} \rangle} &=&0.1\,\textrm{pb}\times \frac{32\pi \langle\xi\rangle^4_{}}{13311 m_{\chi}^2}\nonumber\\
&=&0.1\,\textrm{pb}\times \frac{32\pi \langle\xi\rangle^2_{}}{13311 y_{\chi}^2} \,.
\end{eqnarray}

Due to Eq. (\ref{y34}), the VEV $\langle\xi\rangle$ should have an upper bound,
\begin{eqnarray}
\langle\xi\rangle &\simeq& \left(\frac{13311 y_{\chi}^2 \Omega_\chi^{} h^2}{32 \pi \times  0.1\,\textrm{pb}} \right)^{\frac{1}{2}}_{}\nonumber\\
&=& 97 \,\textrm{TeV}\left(\frac{y_{\chi}^{}}{\sqrt{8\pi/153}}\right) \left(\frac{\Omega_\chi^{} h^2}{0.11}\right)^{\frac{1}{2}}_{}\nonumber\\
&<&97\,\textrm{TeV}\left(\frac{\Omega_\chi^{} h^2}{0.11}\right)^{\frac{1}{2}}_{}\,,
\end{eqnarray}
besides the experimental limit (\ref{low}). The dark matter mass, 
\begin{eqnarray}
\label{dm1}
m_\chi^{} &\simeq &\left(0.1\,\textrm{pb}\times \frac{32\pi\langle\xi\rangle^4_{}}{13311 \Omega_\chi^{} h^2} \right)^{\frac{1}{2}}_{} \nonumber\\
&=& 5\,\textrm{TeV}\left(\frac{\langle \xi\rangle }{34.5\,\textrm{TeV}}\right)^2_{}\left(\frac{0.11}{\Omega_\chi^{} h^2}\right)^{\frac{1}{2}}_{} \,,
\end{eqnarray}
thus should be in the range, 
\begin{eqnarray}
\label{dm2}
&&420\,\textrm{GeV}\left(\frac{0.11}{\Omega_\chi^{} h^2}\right)^{\frac{1}{2}}_{}\lesssim
m_\chi^{} < 40\,\textrm{TeV}\left(\frac{0.11}{\Omega_\chi^{} h^2}\right)^{\frac{1}{2}}_{}\nonumber\\
&&\textrm{for}~~10\,\textrm{TeV} \lesssim \langle\xi\rangle < 97\,\textrm{TeV}\,.
\end{eqnarray}

The gauge interactions can also result in a dark matter scattering off nucleons. The spin-independent cross section is given by \cite{jkg1996}
\begin{eqnarray}
\sigma_{\chi N}^{}&=& \frac{153 g_{B-L}^4}{16 \pi} \frac{\mu_r^2 }{M_{Z_{B-L}^{}}^4} \nonumber\\
&=& \frac{153 }{4\pi} \frac{\mu_r^2 }{\langle\xi\rangle^4_{}}= \frac{8}{87} \frac{\mu_r^2 }{m_\chi^2} \frac{0.1\,\textrm{pb}}{\Omega_\chi^{} h^2_{} }\,.
\end{eqnarray}
Here $\mu_r^{}=m_N^{}m_\chi^{}/(m_N^{}+m_\chi^{})$ is a reduced mass with $m_N^{}$ being the nucleon mass. As the dark matter is much heavier than the nucleon, we can simply read 
\begin{eqnarray}
\sigma_{\chi N}^{}&=& 3\times 10^{-45}_{}\,\textrm{cm}^2_{}\left(\frac{\mu_r^{}}{940\,\textrm{MeV}}\right)^2_{} \left(\frac{0.11}{\Omega_\chi^{} h^2_{}}\right)\nonumber\\
&&\times \left( \frac{5\,\textrm{TeV}}{m_\chi^{}}\right)^2_{} \,.
\end{eqnarray}
To satisfy the dark matter direct detection results \cite{cui2017,aprile2018}, the dark matter mass should have a low limit,
\begin{eqnarray}
\label{dm3}
m_\chi^{} \gtrsim 5\,\textrm{TeV}\,.
\end{eqnarray}
Therefore, the allowed dark matter mass should be
\begin{eqnarray}
\label{range}
&&5\,\textrm{TeV}\left(\frac{0.11}{\Omega_\chi^{} h^2}\right)^{\frac{1}{2}}_{}\lesssim
m_\chi^{} < 40\,\textrm{TeV}\left(\frac{0.11}{\Omega_\chi^{} h^2}\right)^{\frac{1}{2}}_{}\nonumber\\
&&\textrm{for}~~34.5\,\textrm{TeV} \lesssim \langle\xi\rangle < 97\,\textrm{TeV}\,.
\end{eqnarray}

The fermion $\zeta_R^{}$ is massless so that it will affect the effective neutrino number which is stringently constrained by the BBN. We hope the massless $Z_R^{}$ can decouple above the QCD scale and hence give a negligible contribution to the effective neutrino number. For this purpose, we need consider the annihilations of the $\zeta_R^{}$ fermion into the light species,   
\begin{eqnarray}
\sigma_{\zeta}^{} &=&\sum_{f=d,u,s,e,\mu,\nu_L^{}}^{}\sigma(\zeta_R^{}+\zeta_R^c\rightarrow f+f^c) \nonumber\\
&=& \frac{75}{32\pi}\frac{g_{B-L}^4}{M_{Z_{B-L}}^4}s =  \frac{75}{8\pi}\frac{s}{\langle\xi\rangle^4_{}}\,,
\end{eqnarray}
with $s$ being the Mandelstam variable. The interaction rate then should be \cite{gnrrs2003}
\begin{eqnarray}
\Gamma_{\zeta}^{} &=&\frac{\frac{T}{32\pi^4_{}}\int^{\infty}_{0} s^{3/2}_{} K_1^{}\left(\frac{\sqrt{s}}{T}\right) \sigma_{\zeta}^{}ds }{\frac{2}{\pi^2_{}}T^3_{}}= \frac{18}{\pi^3_{}} \frac{T^5_{}}{\langle\xi\rangle^4_{}}\,,
\end{eqnarray}
with $K_1^{}$ being a Bessel function. We take $g_\ast^{}(300\,\textrm{MeV})\simeq 59.75$ and then find 
\begin{eqnarray}
\left[\Gamma_{\zeta}^{} < H(T)\right]_{T\gtrsim 300\,\textrm{MeV}}^{}~~\textrm{for}~~\langle\xi\rangle \gtrsim 37\,\textrm{TeV}\,.
\end{eqnarray}
So, if the massless $\zeta_R^{}$ is harmless, the dark matter mass (\ref{range}) should be modified, i.e.
\begin{eqnarray}
&&5.7\,\textrm{TeV}\left(\frac{0.11}{\Omega_\chi^{} h^2}\right)^{\frac{1}{2}}_{}\lesssim
m_\chi^{} < 40\,\textrm{TeV}\left(\frac{0.11}{\Omega_\chi^{} h^2}\right)^{\frac{1}{2}}_{}\nonumber\\
&&\textrm{for}~~37\,\textrm{TeV} \lesssim \langle\xi\rangle < 97\,\textrm{TeV}\,.
\end{eqnarray}

\section{Conclusion}

In this paper, we have simultaneously explained the neutrino mass, the baryon asymmetry and the dark matter by resorting to a $U(1)_{B-L}^{}$ gauge symmetry. In order to cancel the gauge anomalies, we introduce seven neutral fermions including two usual right-handed neutrinos. After a Higgs singlet develops its VEV for spontaneously breaking the $U(1)_{B-L}^{}$ symmetry, two neutral fermions can obtain their small Majorana masses by integrating out some heavy scalar or fermion singlets, meanwhile, they can have a sizable mass term with the right-handed neutrinos. After the SM Higgs doublet drives the electroweak symmetry breaking, the two right-handed neutrinos and the three left-handed neutrinos can acquire their Dirac masses. We thus can realize an inverse seesaw to give a neutrino mass matrix with two nonzero eigenvalues. Within this framework, the heavy scalar or fermion singlet decays can generate the cosmic baryon asymmetry in association with the sphaleron processes. On the other hand, another two neutral fermions can form a stable Dirac particle due to their Yukawa coupling with the $U(1)_{B-L}^{}$ Higgs singlet. This Dirac fermion can serve as a dark matter particle. As for the seventh neutral fermion, it keeps massless but decouples early.

\textbf{Acknowledgement}: This work was supported by the National Natural Science Foundation of China under Grant No. 11675100 and the Recruitment Program for Young Professionals under Grant No. 15Z127060004.

\appendix

\section{The $U(1)_{B-L}^{}$ gauge anomalies }

The $SU(3)_c^{}-SU(3)_c^{}-U(1)_{B-L}^{}$ anomaly is
\begin{eqnarray}
3\times 3\times \left[2\times \left(+\frac{1}{3}\right) -\left(+ \frac{1}{3}\right) - \left(+\frac{1}{3}\right) \right]=0\,.
\end{eqnarray}

The $SU(2)_L^{}-SU(2)_L^{}-U(1)_{B-L}^{}$ anomaly is 
\begin{eqnarray}
3\times 2 \times  \left[3\times\left(+ \frac{1}{3}\right) +\left(-1\right)\right]=0\,.
\end{eqnarray}

The $U(1)_Y^{}-U(1)_Y^{}-U(1)_{B-L}^{}$ anomaly is 
\begin{eqnarray}
\!\!\!\!&&3\times \left\{3\times \left[2\times \left(+\frac{1}{6}\right)^{\!2}_{} - \left(-\frac{1}{3}\right)^{\!2}_{} - \left(+\frac{2}{3}\right)^{\!2}_{}\right] \times \left(+ \frac{1}{3}\right) \right.\nonumber\\
[2mm]
\!\!\!\!&&\left.+\left[2\times \left(-\frac{1}{2}\right)^2_{} - \left(-1\right)^2_{} \right]\times \left(-1\right) \right\}  =0\,.
\end{eqnarray}

The $U(1)_Y^{}-U(1)_{B-L}^{}-U(1)_{B-L}^{}$ anomaly is 
\begin{eqnarray}
&&3\times \left\{3\times \left[2\times \left(+\frac{1}{6}\right) -\left(-\frac{1}{3}\right)-\left(+\frac{2}{3}\right)\right]\times \left(+\frac{1}{3}\right)^{\!2}_{} \right.\nonumber\\
[2mm]
&&\left.+\left[2\times \left(-\frac{1}{2}\right) -\left(-1\right)\right]\times \left(-1\right) \right\}=0\,.
\end{eqnarray}

The $U(1)_{B-L}^{}-U(1)_{B-L}^{}-U(1)_{B-L}^{}$ anomaly is 
\begin{eqnarray}
\!\!\!\!\!\!\!\!&&3\times \left\{3\times \left[2\times \left(+\frac{1}{3}\right)^{\!3}_{}-\left(+\frac{1}{3}\right)^{\!3}_{}-\left(+\frac{1}{3}\right)^{\!3}_{}\right] \right.\nonumber\\
[2mm]
\!\!\!\!\!\!\!\!&&\left.+\left[2\times \left(-1\right)^{3}_{} -\left(-1\right)^{3}_{}\right] \right\} -2\times \left(-1\right)^3_{} -2\times \left(+\frac{1}{2}\right)^3_{} \nonumber\\
[2mm]
\!\!\!\!\!\!\!\!&&-\left(\frac{1-\sqrt{153}}{4}\right)^3_{} -\left(\frac{1+\sqrt{153}}{4}\right)^3_{} - \left(-\frac{5}{2}\right)^3_{}=0\,.
\end{eqnarray}

The graviton-graviton-$U(1)_{B-L}^{}$ anomaly is
\begin{eqnarray}
\!\!\!\!&&3\times \left\{3\times \left[2\times \left(+\frac{1}{3}\right)-\left(+\frac{1}{3}\right)-\left(+\frac{1}{3}\right)\right] \right.\nonumber\\
[2mm]
\!\!\!\!&&\left.+\left[2\times \left(-1\right) -\left(-1\right)\right] \right\} - -2\times \left(-1\right) -2\times \left(+\frac{1}{2}\right) \nonumber\\
[2mm]
\!\!\!\!&&-\left(\frac{1-\sqrt{153}}{4}\right) -\left(\frac{1+\sqrt{153}}{4}\right)- \left(-\frac{5}{2}\right)=0\,.
\end{eqnarray}


\begin{thebibliography}{99}



 
\bibitem{pdg2018}
M. Tanabashi {\it et al.}, (Particle Data Group), Phys. Rev. D \textbf{98}, 030001 (2018).





\bibitem{minkowski1977}
P. Minkowski, Phys. Lett. B \textbf{67}, 421 (1977).

\bibitem{yanagida1979}
T. Yanagida, {\it Proceedings of the Workshop on Unified Theory and the Baryon Number of the Universe}, ed. O. Sawada and A. Sugamoto (Tsukuba 1979).

\bibitem{grs1979}
M. Gell-Mann, P. Ramond, and R. Slansky, {\it Supergravity}, ed. F. van Nieuwenhuizen and D. Freedman
(North Holland 1979).

\bibitem{ms1980}
R.N. Mohapatra and G. Senjanovi\'{c}, Phys. Rev. Lett. \textbf{44}, 912 (1980).






\bibitem{mw1980}
M. Magg and C. Wetterich, Phys. Lett. B \textbf{94}, 61 (1980).

\bibitem{sv1980}
J. Schechter and J.W.F. Valle, Phys. Rev. D \textbf{22}, 2227 (1980).


\bibitem{cl1980}

T.P. Cheng and L.F. Li, Phys. Rev. D \textbf{22}, 2860 (1980). 

\bibitem{lsw1981}
G. Lazarides, Q. Shafi, and C. Wetterich, Nucl. Phys. B \textbf{181},
287 (1981).

\bibitem{ms1981}
R.N. Mohapatra and G. Senjanovi\'{c}, Phys. Rev. D
\textbf{23}, 165 (1981).





\bibitem{flhj1989}
R. Foot, H. Lew, X.G. He, and G.C. Joshi, Z. Phys. C \textbf{44},
441 (1989).





\bibitem{fy1986}
M. Fukugita and T. Yanagida, Phys. Lett. B \textbf{174}, 45 (1986).



\bibitem{lpy1986}
P. Langacker, R.D. Peccei, and T. Yanagida, Mod. Phys. Lett. A
\textbf{1}, 541 (1986); M.A. Luty, Phys. Rev. D \textbf{45}, 455
(1992); R.N. Mohapatra and X. Zhang, Phys. Rev. D \textbf{46}, 5331 (1992).





\bibitem{fps1995}
M. Flanz, E.A. Paschos, and U. Sarkar, Phys. Lett. B \textbf{345},
248 (1995); M. Flanz, E.A. Paschos, U. Sarkar, and J. Weiss, Phys.
Lett. B \textbf{389}, 693 (1996); L. Covi, E. Roulet, and F.
Vissani, Phys. Lett. B \textbf{384}, 169 (1996); A. Pilaftsis, Phys.
Rev. D \textbf{56}, 5431 (1997).


\bibitem{ms1998}
E. Ma and U. Sarkar, Phys. Rev. Lett. \textbf{80}, 5716 (1998).


\bibitem{bcst1999}
R. Barbieri, P. Creminelli, A. Strumia, and N. Tetradis, Nucl. Phys. B \textbf{575}, 61 (2000).




\bibitem{hambye2001}
T. Hambye, Nucl. Phys. B \textbf{633}, 171 (2002).


\bibitem{di2002}
S. Davidson and A. Ibarra, Phys. Lett. B \textbf{535}, 25 (2002); W.
Buchm\"{u}ller, P. Di Bari, and M. Pl\"{u}macher, Nucl. Phys. B
\textbf{665}, 445 (2003).

\bibitem{gnrrs2003}
G.F. Giudice, A. Notari, M. Raidal, A. Riotto, and A. Strumia, Nucl. Phys. B \textbf{685}, 89 (2004).


\bibitem{hs2004}
T. Hambye and G. Senjanovi\'{c}, Phys. Lett. B \textbf{582}, 73
(2004); S. Antusch and S.F. King, Phys. Lett. B \textbf{597}, 199
(2004).


\bibitem{bbp2005}
W. Buchmuller, P. Di Bari, and M. Plumacher, Annals Phys. \textbf{315}, 305 (2005).



\bibitem{krs1985}
V.A. Kuzmin, V.A. Rubakov, and M.E. Shaposhnikov, Phys. Lett. B
\textbf{155}, 36 (1985).



\bibitem{mv1986}
R. Mohapatra and J. W. F. Valle, Phys. Rev. D \textbf{34}, 1642 (1986).






\bibitem{adefhv2018} 
K. Agashe, P. Du, M. Ekhterachian, C.S. Fong, S. Hong, and L. Vecchi, Phys. Lett. B \textbf{785}, 489 (2018); K. Agashe, P. Du, M. Ekhterachian, C.S. Fong, S. Hong, and L. Vecchi, JHEP \textbf{1904}, 029 (2019).


\bibitem{gu2019}
P.H. Gu, arXiv:1907.09444 [hep-ph]. 




\bibitem{rsv2019}
N. Rojas, R. Srivastava, and J.W.F. Valle, arXiv:1907.07728 [hep-ph].














\bibitem{mp2007}
J.C. Montero and V. Pleitez, Phys. Lett. B \textbf{675}, 64 (2009). 


\bibitem{pry2016} 
S. Patra, W. Rodejohann, and C.E. Yaguna, JHEP \textbf{1609}, 076 (2016).









\bibitem{gu2019-2}
P.H. Gu, arXiv:1907.10018 [hep-ph].




\bibitem{ht1990}
J.A. Harvey and M.S. Turner, Phys. Rev. D \textbf{42}, 3344 (1990).




\bibitem{kt1990}
E.W. Kolb and M.S. Turner, \textit{The Early Universe},
Addison-Wesley, 1990.

 
 





\bibitem{gu2019-3}
P.H. Gu, arXiv:1907.10019 [hep-ph].
  
 
 
\bibitem{afpr2017}
S. Alioli, M. Farina, D. Pappadopulo, and J.T. Ruderman, Phys. Rev. Lett. \textbf{120}, 101801 (2018).




\bibitem{klq2016}
M. Klasen, F. Lyonnet, and F.S. Queiroz, Eur. Phys. J. C \textbf{77}, 348 (2017).

 



\bibitem{bhkk2009}
M. Beltran, D. Hooper, E.W. Kolb, and Z.C. Krusberg, Phys. Rev. D \textbf{80}, 043509 (2009);
K. Cheung, P.Y. Tseng, and T.C. Yuan, JCAP \textbf{1101}, 004 (2011).




\bibitem{jkg1996}
G. Jungman, M. Kamionkowski, and K. Griest, Phys. Rept. \textbf{267}, 195 (1996).




\bibitem{cui2017}

X. Cui {\it et. al.}, (PandaX-II Collaboration), Phys. Rev. Lett. \textbf{119}, 181302 (2017). 



\bibitem{aprile2018}

E. Aprile {\it et. al.}, (XENON Collaboration), Phys. Rev. Lett. \textbf{121}, 111302 (2018).



\end{thebibliography}
\end{document}